\documentclass[]{spie}  

\usepackage{amsmath,amsfonts,amssymb}
\usepackage{graphicx}
\usepackage[colorlinks=true, allcolors=blue]{hyperref}

\title{Operational modes and efficiency of SOXS}

\author[a]{Claudi R.}
\author[a,b]{Biondi F.}
\author[a]{Elias-Rosa N.}
\author[c]{Genoni M.}
\author[d]{Munari M.}
\author[a]{Radhakrishnan K.}
\author[a]{Ricci D.}
\author[d]{Zanmar Sanchez R.}
\author[c]{Campana S.}
\author[e]{Schipani P.}
\author[c]{Aliverti M.}
\author[a]{Baruffolo A.}
\author[f,g]{Ben--Ami S.}
\author[h,i]{Brucalassi A.}
\author[e]{Capasso G.}
\author[j,d]{Cosentino R.}
\author[k]{D'Alessio F.}
\author[c]{D'Avanzo P.}
\author[f]{Hershko O.}
\author[l,m]{Kuncarayakti H.}
\author[c]{Landoni M.}
\author[h,o]{Pignata G.}
\author[p]{Rubin A.}
\author[d,q]{Scuderi S.}
\author[k]{Vitali F.}
\author[r]{Young D.}
\author[s]{Achr\'en J.}
\author[t,o]{Araiza--Duran J.A.}
\author[u]{Arcavi I.}
\author[f]{Bruch}
\author[a]{Cappellaro E.}
\author[e]{Colapietro M.}
\author[e]{Della Valle M.}
\author[d]{Di Benedetto R.}
\author[a]{De Pascale M.}
\author[e]{D'Orsi S.}
\author[j]{Hernandez M.}
\author[f]{Gal--Yam A.}
\author[v]{Li Causi G.}
\author[a]{Marafatto L.}
\author[l]{Matila S.}
\author[f]{Rappaport M.}
\author[c]{Riva M.}
\author[a]{Salasnich B.}
\author[r]{Smartt S.}
\author[z]{Stritzinger M.}
\author[a]{Turatto M.}
\author[j]{Perez Ventura H.}

\affil[a]{INAF--Osservatorio Astronomico di Padova, vicolo Osservatorio 5, 35122 Padova, Italy}
\affil[b]{Max-Planck-Institut f\"ur Extraterrestrische Physik, Giessenbachstr. 1, D-85748 Garching, Germany }
\affil[c]{INAF-- Osservatorio Astronomico di Brera, via Bianchi 46, 23807 Merate (LC), Italy}
\affil[d]{INAF--Osservatorio Astrofisico di Catania, via di Santa Sofia 78, 95123 Catania, Italy}
\affil[e]{INAF--Osservatorio Astronomico di Capodimonte, salita Moiariello 16, 80131 Napoli, Italy}
\affil[f]{Weizmann Institute of Science, Herzl St 234, Rehovot, 7610001, Israel}
\affil[g]{Harvard Smithsonian Center for Astrophysics, Cambridge, USA}
\affil[h]{Universidad Andres Bello, Avda. Republica 252, Santiago, Chile}
\affil[i]{INAF--Osservatorio Astronomico di Firenze, Largo E. Fermi 5, 50125, Firenze, Italy}
\affil[j]{FGG--INAF, TNG, Rambla J.A. Fern\'andez P\'erez 7, 38712 Bre\~na Baja (TF), Spain}
\affil[k]{INAF--Osservatorio Astronomico di Roma, via Frascati 33, 00078, Monte Porzio Catone (Roma), Italy}
\affil[l]{Tuorla Observatory, Dep. of Physics and Astronomy, 20014 University of Turku, Finland}
\affil[m]{Finnish Centre for Astronomy with ESO (FINCA), 20014 University of Turku, Finland}
\affil[n]{Universidad Andres Bello, Avda. Republica 252, Santiago, Chile}
\affil[o]{Millennium Institute of Astrophysics (MAS), Nuncio Monsenor Sotero Sanz 100, Providencia, Santiago, Chile}
\affil[p]{ESO, Karl Schwarzschild Strasse 2, D--85748, Garching bei M\"unchen, Germany}
\affil[q]{INAF - Istituto di Astrofisica Spaziale e Fisica Cosmica, Via Corti 12, I-20133 Milano, Italy}
\affil[r]{Astrophysics Research Centre, Queen's University, Belfast, County Antrim, BT7 1NN, UK}
\affil[s]{Incident Angle Oy, Capsiankatu 4 A 29, 20320 Turku, Finland}
\affil[t]{Centro de Investigaciones en Optica A. C., Loma del Bosque 115, Lomas del Campestre, 37150 Leon Guanajuato, Mexico}
\affil[u]{Tel Aviv University, Department of Astrophysics, 69978 Tel Aviv, Israel}
\affil[v]{INAF - Istituto di Astrofisica e Planetologia Spaziali, Via Fosso del Cavaliere, I-00133 Roma}
\affil[z]{Aarhus University, Ny Munkegade 120, D-8000, Denmark}

\authorinfo{Further author information: (Send correspondence to R. Claudi)\\R. Claudi: E-mail: riccardo.claudi@inaf.it}

\begin{document} 
\maketitle

\begin{abstract}
Son of X-Shooter (SOXS) will be a high-efficiency spectrograph with a mean Resolution-Slit product of $\sim 4500$ over the entire band capable of simultaneously observing the complete spectral range 350-2000 nm. It consists of three scientific arms (the UV-VIS Spectrograph, the NIR Spectrograph, and the Acquisition Camera) connected by the Common Path system to the NTT, and the Calibration Unit. The Common Path is the backbone of the instrument and the interface to the NTT Nasmyth focus flange. The instrument project went through the Final Design Review in 2018 and is currently in Assembly Integration and test (AIT) Phase. This paper outlines the observing modes of SOXS and the efficiency of each subsystem and the laboratory test plan to evaluate it. 
\end{abstract}


\section{INTRODUCTION}
\label{sec:intro}  
The research on transients has expanded significantly in the past two decades, leading to some of the most recognized discoveries in astrophysics (e.g. gravitational wave events, gamma-ray bursts, super-luminous supernovae, accelerating universe). Nevertheless, so far most of the transient discoveries still lack an adequate spectroscopic follow-up. Thus, it is generally acknowledged that with the availability of so many transient imaging surveys in the next future, the scientific bottleneck is the spectroscopic follow-up of transients. Within this context, SOXS aims to significantly contribute bridging this gap. It will be one of the few spectrographs on a dedicated telescope with a significant amount of observing time to characterize astrophysical transients. It is based on the concept of X-Shooter \cite{vernetetal2011} at the VLT but, unlike its “father”, the SOXS science case is heavily focused on transient events. Foremost, it will contribute to the classifications of transients, i.e. supernovae, electromagnetic counterparts of gravitational wave events, neutrino events, tidal disruptions of stars in the gravitational field of supermassive black holes, gamma-ray bursts and fast radio bursts, X-ray binaries and novae, magnetars, but also asteroids and comets, activity in young stellar objects, and blazars and AGN.

\begin{figure} [ht]
   \begin{center}
   \begin{tabular}{c} 
   \includegraphics[height=10cm]{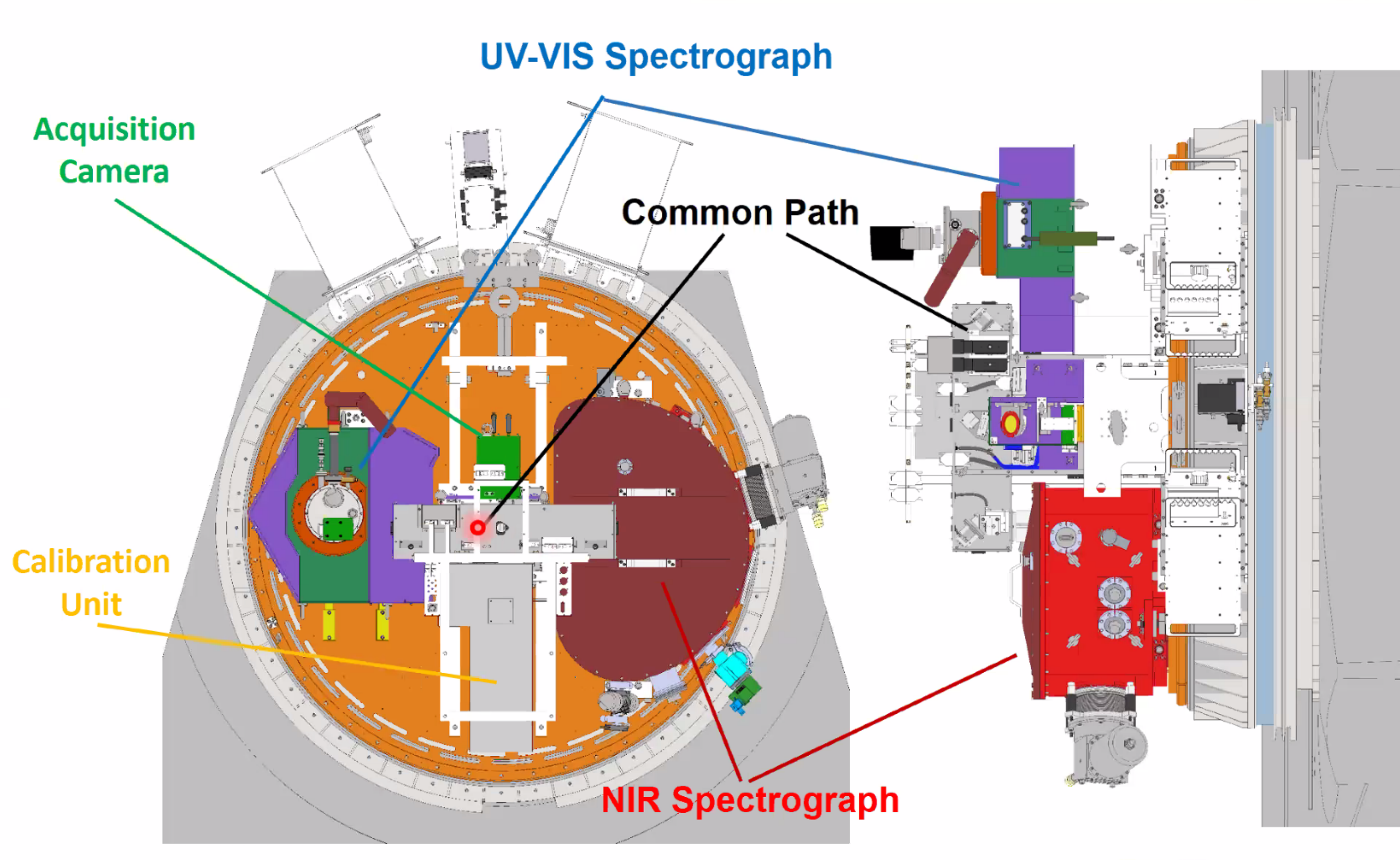}
   \end{tabular}
   \end{center}
   \caption[example] 
   { \label{fig:soxs1} 
SOXS: front and side view of the instrument with the identification of all the sub-systems.}
   \end{figure}

SOXS\cite{schipanietal2018spie, schipanietal2016spie, claudietal2018frap, schipanietal2020spie} will simultaneously cover the electromagnetic spectrum from 0.35 to 2.0\ $\mu$m using two arms (UV--VIS and NIR) with a product slit--resolution of $\sim 4500$. The throughput should enable to reach a S/N$\sim 10$ in a 1-hour exposure of an R=20 mag point source. SOXS, that will see its first light at the end of 2021, will be mounted at the Nasmyth focus of NTT replacing SOFI. The whole system (see Figure\ \ref{fig:soxs1}) is constituted by the three main scientific arms: the UV--VIS spectrograph\cite{rubinetal2018spie,rubinetal2020spie,  cosentinoetal2018spie,cosentinoetal2020spie}, the NIR Spectrograph\cite{vitalietal2018spie, vitalietal2020spie} and the acquisition camera (AC)\cite{brucalassietal2018spie, brucalassietal2020spie}. The three main arms, the calibration box\cite{kuncarayaktietal2020spie} and the NTT are connected together\cite{claudietal2018spie, biondietal2018spie, biondietal2020spie} by the Common Path (CP). 

The main characteristics of the three scientific arms are listed in Table\ \ref{tab:subsy}.

\begin{table}[ht]
\caption{Main characteristics of the SOXS sub systems connected to the SOXS common path.} 
\label{tab:subsy}
\begin{center}       
\begin{tabular}{|l|c|c|c|} 
\hline
\rule[-1ex]{0pt}{3.5ex}        & \textbf{AG Camera} & \textbf{UV--VIS}& \textbf{NIR}  \\
\hline
\rule[-1ex]{0pt}{3.5ex}  F/\# & 3.6 & 6.5 & 6.5   \\
\hline
\rule[-1ex]{0pt}{3.5ex}  Spectral Range& ugrizY $+$ V& 350--850 nm& 800 -- 2000 nm  \\
\hline
\rule[-1ex]{0pt}{3.5ex}  Resolution& $-$ & $>3600\ \sim4500$\ Avg & 5000  \\
\hline
\rule[-1ex]{0pt}{3.5ex}  Slit Width (arcsec)& $-$ & \multicolumn{2}{|c|}{$0.5 - 1.0 -1.5-5.0$}  \\
\hline
\rule[-1ex]{0pt}{3.5ex}  Slit height (arcsec)& $-$ & \multicolumn{2}{|c|}{$12.0$} \\
\hline
\rule[-1ex]{0pt}{3.5ex}  Pixel Scale (arcsec/px)& 0.205 & 0.280 & 0.164  \\
\hline 
\rule[-1ex]{0pt}{3.5ex}  Detector & Andor Ikon M-934 1k $\times$ 1k& e2v CCD44--82 2k$\times$ 4k& H2RG\ 2k $\times$ 2k  \\
\hline 
\rule[-1ex]{0pt}{3.5ex}  Pixel Size ($\mu$m) & 13.0& 15.0& 18.0  \\
\hline 
\end{tabular}
\end{center}
\end{table}

\section{SOXS Operational Modes}
\label{sec:som}  
Once on the mountain, SOXS will be used in three well distinct operational mode:
\begin{itemize}
    \item science mode;
    \item calibration mode;
    \item maintenance mode.
\end{itemize}
The science mode is the usual way in which the whole instrument will be used. The calibration mode collects all those operations that will allow us to remove the instrumental signatures from the scientific data i.e. translate the raw spectra into physical units (wavelength, spatial scale and flux). The maintenance mode is more a status than a real mode in which the instrument is posed in order to allow for operators to perform maintenance operations. In the following we will describe each mode separately.

   \begin{figure} [ht]
   \begin{center}
   \begin{tabular}{c} 
   \includegraphics[height=10cm]{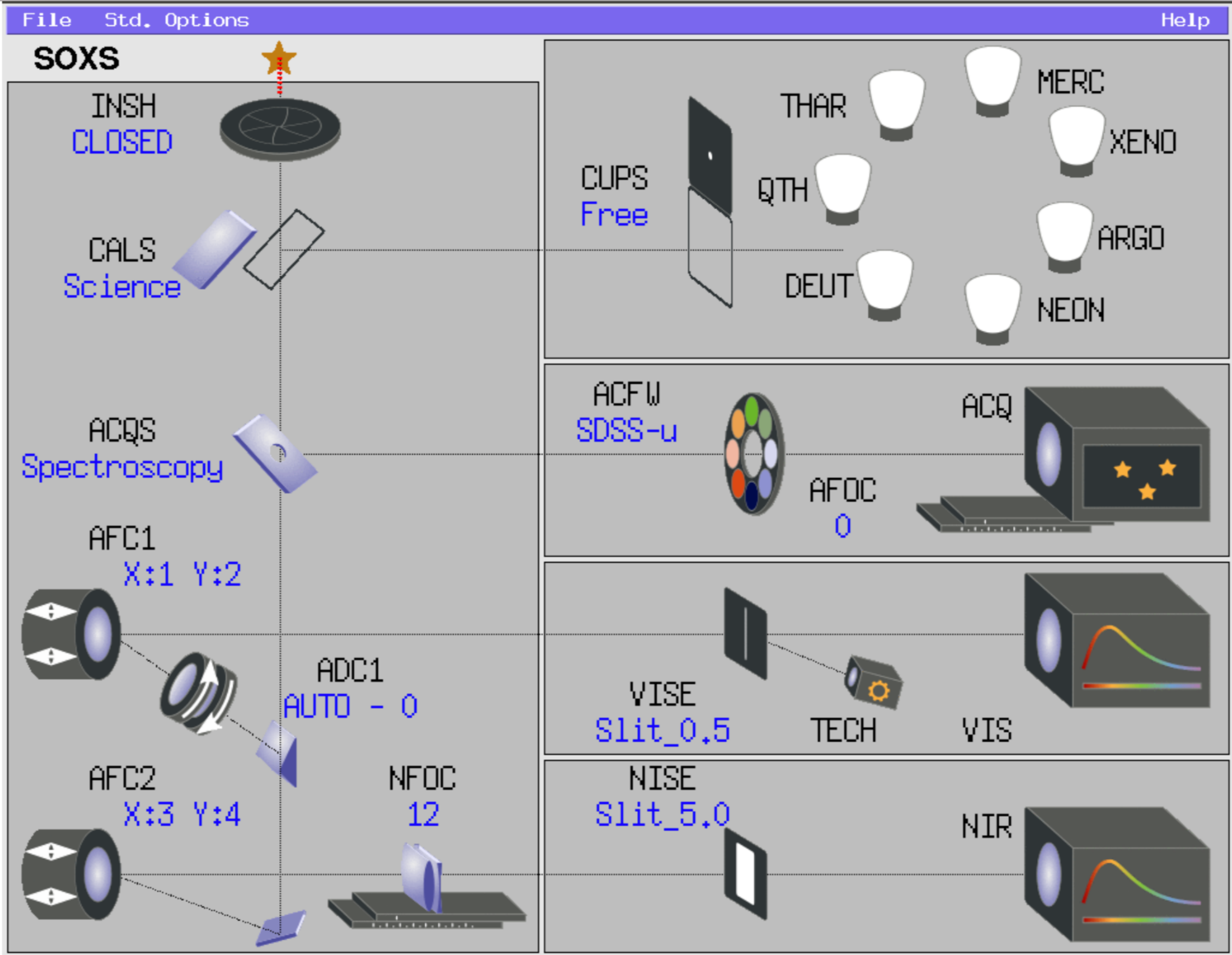}
   \end{tabular}
   \end{center}
   \caption[example] 
   { \label{fig:mode} 
The operational scheme of SOXS as it appears on the synoptic screen of the instrument control system \cite{soxsricci}.}
   \end{figure} 

\subsection{Science mode}
\label{sec:scimo}  

SOXS will be offered for Service Observing, in all of its Observing Modes. Following the science case, SOXS will be a \textit{point and shoot} instrument. This implies that the number of observing modes that may be selected by the user shall be as small as possible to simplify the operation, calibration, pipeline data reduction, and data quality control of SOXS. All of these tasks will benefit from having a simple instrument with few modes; also the reliability will be enhanced.
The science observing modes that are foreseen are: 
\begin{enumerate}
    \item Standard Slit Spectroscopy: in this mode, the observer can use both the spectrographs UV--VIS and NIR (see Figure\ \ref{fig:mode}). This will be the usual observing mode. In this mode it will be not possible to have  contemporaneously the AC on line performing also photometry. Photometry of targets will be performed only in the light imaging mode. The guiding system of the NTT will perform the pursuit of the observed star by means its own guiding camera. In same occasions, it will be possible to select the pellicle position of the acquisition slide into the CP (see Figure\ \ref{fig:mode}), to have also the AC on line together with the two spectrographs for secondary guide.   
    \item Light Imaging: In this mode the user will carry out the photometric observation of the target. The photometric band are listed in Table\ \ref{tab:subsy}. The imaging mode is activated by introducing in the optical path the AC by means of the acquisition slide (see Figure\ \ref{fig:mode})
\end{enumerate}
These modes shall correspond to a single instrument setting that is fully defined just by the mode name, with only a short list of selectable slit widths. The detectors are used in staring mode with a standard setting for read speed and binning; the resulting data format is fixed.
The hardware design shall also be such that the number of user-selectable options that are left open by the design is minimized, in a way that is compatible with the main scientific goal of the instrument.
The following sub-modes shall be implemented up to the commissioning at the telescope, with the decision on the modalities with which they are offered taken by ESO: {\it a)} CCD binning factors; {\it b)} CCD detector read out speeds; {\it c)} CCD read-out window.

\subsection{Calibration Mode}
\label{sec:calmo}  
The goal of the calibrations is to remove the instrumental signatures from the scientific data, translating the raw spectra into physical units (wavelength, spatial scale, and flux). To allow the calibration, the calibration mirror is inserted in the beam by the calibration slide inside the CP (see Figure\ \ref{fig:mode}). The mirror will deflect the light coming from the several lamps (spectral and white lamps) inside the calibration box (see Figure\ \ref{fig:mode}). The observer, setting the observation block before the observation, will select the necessary spectral lamps for the wavelength calibration. Instead, as dark frames and flat fields concern, the instrumental software will perform them in an automatic way. For the dark frames, the system will analyze the several exposure time used during the nights, and it will perform the dark frames exploiting those exposure times. The system will perform automatically the sky flats in the morning just after the observation. Furthermore, for quick reduction purposes, a full set of calibration with all setups will be taken on a weekly basis.

\subsection{Maintenance Mode}
\label{sec:mainmode}  
During its life, SOXS will overhaul to fix some inconvenience or to substitute any worn piece or element. After the substitution, before to restart the SOXS working operations, the operators can check the instrument status. To do this check, it will perform the template of self test (see \cite{ricci2018spie} for details) excluding the connection with the TCS (telescope Control Software).

\section{Efficiency of SOXS}
\label{sec:som}  
\begin{figure} [ht]
   \begin{center}
   \begin{tabular}{c} 
   \includegraphics[height=11cm]{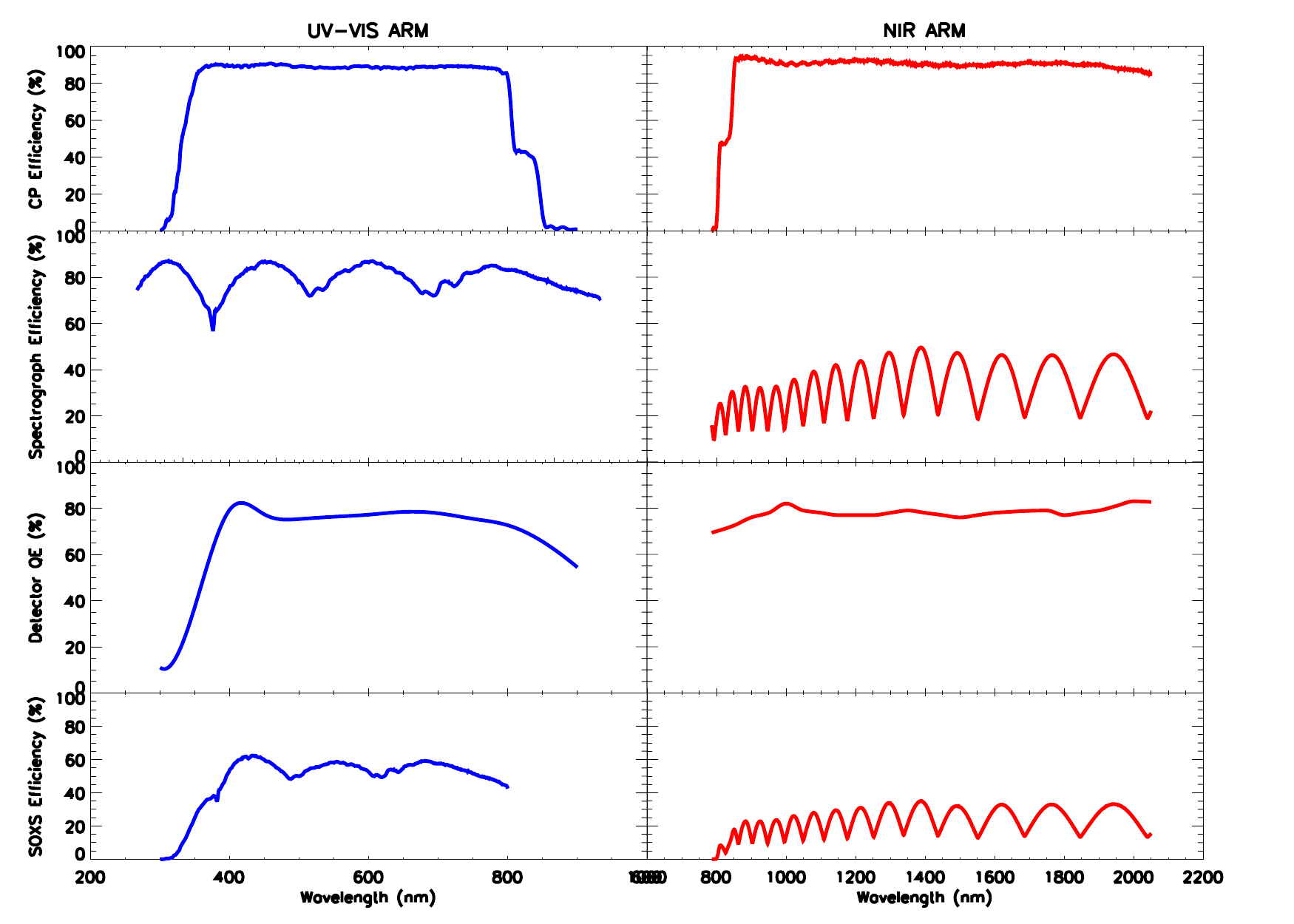}
   \end{tabular}
   \end{center}
   \caption[example] 
   { \label{fig:eff} 
The efficiency of SOXS. \textbf{Left}: from top to bottom, the efficiency of the uv-vis arm of Common path, the UV-VIS Spectrograph, the QE of the CCD, and the total efficiency of the UV-VIS arm. \textbf{Right}: from top to bottom the efficiency of the NIR arm of Common path, the NIR Spectrograph, the QE of the Hawaii - 2RG Detector, and the total efficiency of the NIR arm. Both total efficiency curve do not consider the efficiency of the telescope.}
   \end{figure}

Due to the SOXS aims, the whole instrument shall be efficient, and the optics have been drawn to have efficiency as high as possible. In fact, the spectrograph has to cope with the duty of pursuing transient alarms and characterize faint sources. As described in Section\ \ref{sec:scimo}, the main observing mode of SOXS will be the spectroscopical one. The light coming from the NTT is injected into the CP.  The CP distributes the light towards the two spectrographs that imaged the spectra on their detectors. An e2v CCD44-82 for the UV-VIS (2k$\times$4k) and a Hawaii-2RG (1k $\times$ 1k) for the NIR  spectrograph. Figure\ \ref{fig:eff} shows the efficiency for the two arms (UV-VIS and NIR) of the CP and of each sub-system in cascade. The bottom panel of Figure\ \ref{fig:eff} shows the total efficiency of the two arms. The atmosphere and both the efficiency of the slits and of the telescope are not considered.

\begin{figure} [ht]
   \begin{center}
   \begin{tabular}{c} 
   \includegraphics[height=11cm]{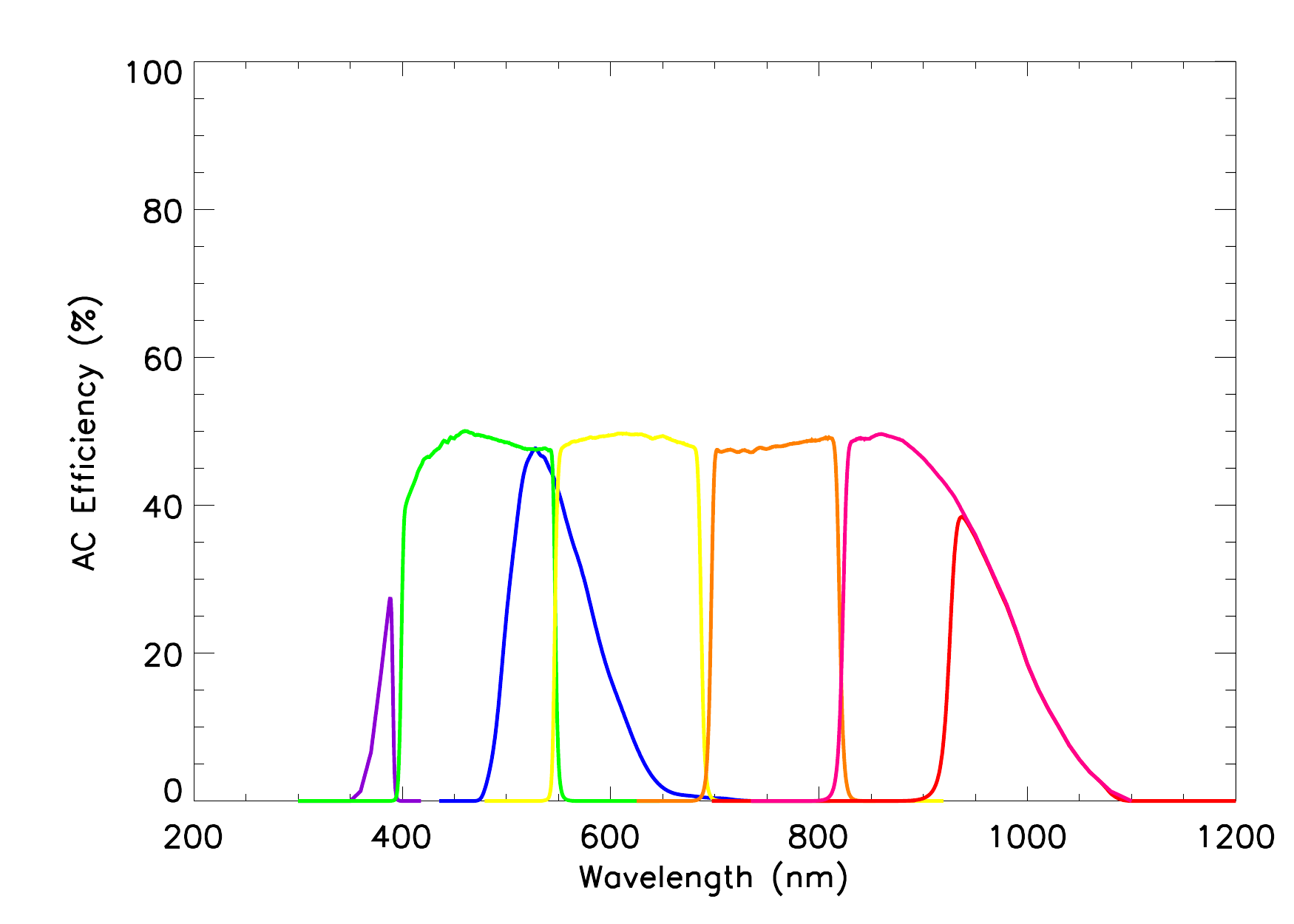}
   \end{tabular}
   \end{center}
   \caption[example] 
   { \label{fig:aceff} 
The AC efficiency in each band as defined by each filter (\textit{u, v, g, r, i, z}, and Y that are in the AC filter wheel.}
   \end{figure}

In the case of photometric mode, the light coming by the telescope passes through the CP and is reflected towards the acquisition camera through a flat mirror on the acquisition slide (see Figure\ \ref{fig:mode}). The acquisition Camera has an optical relay to fit the numerical aperture of the Nasmyth focus on the detector. In this way, the camera has an FoV of $3.5 \times 3.5\ \text{arcmin}^2$ and a pixel scale of $0.205\ \text{arcsec/pixel}$. The AC Detector is an Andor iKon M934 with a pixel size of $13.0\ \mu$m. Before the AC, there is a filter wheel with \textit{u, v, g, r, i}, Y, and Z filters. Figure\ \ref{fig:aceff} shows the efficiency of the whole system, but the telescope and the atmosphere.  All the bands have been checked with the Sloan and LSST bands, and they all fit well. \textit{u} band is narrower than the corresponding  SLOAN \textit{u} band. This is due to the Andor CCD QE that has been optimized for the red. Details on the AC, and its set of filters, and their performances are given in Brucalassi et al. (2020)\cite{brucalassietal2020spie}.

The total efficiency will allow SOXS to compete with the brand new spectrographs that are or will be mounted on telescopes of 4 m class or higher. For example, SOXS can compete with the low-resolution spectrograph \textit{Next Generation Palomar Spectrograph} (NGPS) that is under construction for the Palomar telescope (D$\sim 5$\ m). The claimed efficiency (from slit to the detector) of NGPS is 80\% (see Figure 5 of Kulkarni, 2020 \cite{kulkarni2020arxiv}).
\section{Tests}
The SOXS efficiency will be tested both in the laboratory and at the telescope.  
The laboratory test will be performed on CP using photodiodes or pyroelectric energy sensors.  In this way, we will test the CP to be sure that the efficiency in both the UV-VIS and NIR arms are those that we are expecting. The test will be performed on a selected set of wavelengths (see Table\ \ref{tab:nir}). These wavelengths have been chosen because they identify particular zones of the imaged spectra. They are central wavelengths of the pseudo orders of the UV-VIS Spectrograph and the blaze wavelengths of the NIR Spectrograph. Moreover, we also test the wavelengths at the two limits of each UV-VIS pseudo order and of NIR echelle order. 

\begin{table}[h]
\caption{List of wavelength for each pseudo order (UV-VIS Spectrograph) and echelle order (NIR Spectrograph) to be used to test the performances of UV-VIS in laboratory.}
\begin{center}
\begin{tabular}{cccc}
\hline
Ord. & $\lambda_1$ &$\lambda_2$ &$\lambda_3$ \\
       &    (nm)           &   (nm)           &   (nm)           \\
\hline
\multicolumn{4}{c}{UV-VIS Spectrograph }\\
\hline
$u$    &360      &390    &420\\
$g$    & 445     &485    & 525\\
$r$    & 550     & 600   & 660\\
$i$      & 690    & 760    & 830\\
\hline
\hline
\multicolumn{4}{c}{NIR Spectrograph } \\
\hline
9     & 2.025 & 2.144& 2.263\\
10  &  1.833 & 1.930 & 2.026\\
11  &  1.674 & 1.754 &1.834 \\
12  &  1.541 & 1.608 &1.608\\
13  &  1.427 & 1.484 & 1.541 \\
14  &  1.329 & 1.378 & 1.428\\
15  &  1.244 & 1.286 & 1.329\\
16  &  1.168 & 1.206 &1.244\\
17  &  1.102 & 1.135 & 1.168\\
18  &  1.042 & 1.072 &1.102\\
19  &  0.989 & 1.016 &1.042\\
20  &  0.941 & 0.965 & 0.989\\
21  &  0.897 & 0.919 & 0.941\\
22  &  0.857 & 0.877 & 0.897\\
23  &  0.821 & 0.839 & 0.857\\
24  &  0.787 & 0.804 &0.821\\
\hline
\hline
\end{tabular}
\end{center}
\label{tab:nir}
\end{table}%

The laboratory test will be performed in a dark room, where we will check the loss of flux of the source (the telescope simulator) after the insertion of the CP between the light source and the photodiode (on UV‐VIS and NIR arms foci). A unique pyroelectric sensor could replace the photodiodes. 

As the spectrographs efficiency concern, the better test to do will be at the telescope.  After we will mount SOXS on the NTT, we will observe a set of flux calibration stars to be able to evaluate the total efficiency (detector, plus spectrograph, plus telescope and atmosphere) of the system. As an example, we can use the set of stars described in Table\ \ref{tab:flstd}, refereed as the X-shooter Reference Spectra. They are five spectro- photometric standard stars for which exists theoretically X-shooter spectra and are hot white dwarfs (WD) or subdwarfs (sdO) with independent flux information in the wavelength range between $300\ \text{and}\ 2500\ \text{nm}$. Other possibilities are the spectrophotometric standard stars by Oke (1990)\cite{oke1990aj} or also those by Hamuy (1992, 1994)\cite{hamuyetal1992pasp, hamuyetal1994pasp}. These are all stars with flux well defined in the wavelength range between $\sim 330$ and $\sim 1000$\ nm \footnote{https://www.eso.org/sci/observing/tools/standards/spectra.html}.

\begin{table}[ht]
\caption{Spectro-photometric stars in the X-Shooter Reference spectra set.} 
\label{tab:flstd}
\begin{center}       
\begin{tabular}{ccccccccc} 
\hline
Star & alpha & delta& Type & (U-B)& (B-V) & V & PM(RA) & PM(Dec) \\
     &\multicolumn{2}{c}{(Eq. 2000,Epoch.J2000)}&\multicolumn{2}{c}{($"$/yr)} \\
\hline
 GD 71  & 05:52:27.61& +15:53:13.8& DA& $-1.11$ &  $-0.25$ &  13.03 & 0.08& $-0.17$\\
 LTT 3218 & 08:41:32.50& $-$32:56:34.0&  DA&  $-0.55$ &  +0.22&  11.86 &  $-1.10$ &  1.34\\
 EG 274 & 16:23:33.84& $-$39:13:46.2 &  DA  & $-0.97$  &  $-0.14$ &  $11.03$ & $0.08$ &  0.00 \\
 LTT 7987& 20:10:56.85 & $-$30:13:06.6 & DA & $-0.63$ &  +0.04& 12.24&   $-0.34$ & $-0.25$ \\
 Feige 110 & 23:19:58.40 & $-$05:09:56.2& sDO&  $-1.09$ &  $-0.05$ &  11.50 &  $-0.01$ &  0.00 \\
\hline 
\end{tabular}
\end{center}
\end{table}

\section{Conclusion}
\label{sec:conclusion}  
The SOXS instrument will be used to characterize transient sources. To pursue this aim,  it shall be as simple as possible, rapid in response to an alarm, and efficient. We describe the working modes of the instrument and their efficiency (with no slit, telescope, and atmosphere considered) and the way to test it both in the laboratory and in the sky. A list of possible standard stars to be observed to evaluate the whole system efficiency, is also given.

\bibliography{report2020} 
\bibliographystyle{spiebib} 

\end{document}